\documentclass[
structabstract
]{aa}

\usepackage[fleqn]{amsmath}

\usepackage{graphicx}

\usepackage{txfonts}

\usepackage{multirow}

\usepackage{natbib}
\bibpunct{(}{)}{;}{a}{}{,}

\usepackage[pdftex=true,colorlinks=true]{hyperref}
\usepackage[all]{hypcap}
\hypersetup{
pdfpagemode = {UseOutlines},
baseurl = {http://www.mpia.de/homes/kuiper},
pdftitle = {The reliability of approximate radiation transport methods for irradiated disk studies},
pdfauthor = {R.~Kuiper},
pdfkeywords = {radiative transfer - hydrodynamics - accretion disks - stars: circumstellar matter - stars: formation - methods: numerical}
}

\title{The reliability of approximate radiation transport methods\\for irradiated disk studies}

\author{ 
R.~Kuiper\inst{1,2} 
\and R.~S.~Klessen\inst{3} 
}

\institute{
Universit\"at T\"ubingen,
Institut f\"ur Astronomie und Astrophysik,
Computational Physics,
Auf der Morgenstelle 10,
D-72076 T\"ubingen,
Germany
\\
\email{rolf.kuiper@uni-tuebingen.de}
\and 
Max-Planck-Institut f\"ur Astronomie Heidelberg, 
K\"onigstuhl 17, 
D-69117 Heidelberg, 
Germany
\and
Universit\"at Heidelberg,
Zentrum f\"ur Astronomie Heidelberg,
Institut f\"ur theoretische Astrophysik,
Albert-\"Uberle-Strasse 2,
D-69120 Heidelberg, 
Germany
}

\date{ Received {\it date} / Accepted {\it date} }

\abstract{
Dynamical studies of irradiated circumstellar disks require 
an accurate treatment of radiation transport to, for example, properly determine cooling and fragmentation properties.
At the same time the radiation transport algorithm should be as fast as the (magneto-) hydrodynamics to allow for an efficient usage of computing resources.
Such fast radiation transport methods imply the acceptance of far-reaching approximations.
} 
{
We check the reliability of fast, approximate radiation transport methods for circumstellar disk studies by comparing their accuracy to previous standard radiation benchmark test results.
} 
{
We use different approximate radiation transport methods and compute the equilibrium temperature distribution in a setup of a central star and a slightly flared circumstellar disk, which is embedded in an optically thin envelope. 
We perform simulations for a wide range of optical depths of the disk's midplane from $\tau_\mathrm{550nm}=0.1$ up to $\tau_\mathrm{810nm}=1.22\times10^{+6}$.
We check the accuracy of
the gray flux-limited diffusion (FLD) approximation and
the gray and frequency-dependent hybrid approximation.
In the hybrid method, the stellar irradiation is computed via a gray or frequency-dependent ray-tracing (RT) step and the thermal (re-)emission by dust grains is shifted to a gray FLD solver.
} 
{
1. For moderate optical depths, a gray approximation of the stellar irradiation yields a slightly hotter inner rim and a slightly cooler midplane of the disk at larger radii, but is otherwise in agreement with the frequency-dependent treatment.\\
2. The gray FLD approximation fails to compute an appropriate temperature profile in all regimes of optical depth; the maximum deviations to the comparison runs are 50\% in the optically thin and up to 280\% in the optically thick limit.
For low optical depth, the isotropic assumption within the FLD method yields a too steep decrease of the radial temperature slope.
For higher optical depths, the FLD approximation does not reproduce the shadow behind the optically thick inner rim of the circumstellar disk, yielding artificial heating at larger disk radii.\\
3. The frequency-dependent RT + gray FLD approximation yields remarkable accuracy for the whole range of optical depths.
} 
{
The high accuracy of the frequency-dependent hybrid radiation transport algorithm makes this method ideally suited for (magneto-) hydrodynamical studies of irradiated circumstellar disks.
}


\begin{document}

\maketitle

\section{Introduction}
\label{sect:introduction}
The formation, morphology, and evolution of circumstellar disks depend crucially on their temperature distribution.
After their formation, angular momentum transport can be enabled by various instabilities.
Purely hydrodynamic instabilities \citep[e.g.,][]{Lin:1982p16973, Godon:2000p16974, Klahr:2003p779, Umurhan:2004p16978, Johnson:2005p16981},
magneto-rotational instability \citep{Balbus:1991p816, Hawley:1991p643}, and
gravitational instability \citep{Boss:1984p16962}
denote important paths to sustain accretion.
The strength of the angular momentum transport and accretion depends on the temperature distribution of the disk. 
For purely hydrodynamic instabilities, this was shown, for example, in the publication series \citet{Kley:1992p560}, \citet{Kley:1993p656}, and \citet{Kley:1993p480};
for magneto-rotational instability in local shearing box simulations it was shown by \citet{Hirose:2011p17114} and \citet{Flaig:2012p17017}, and
for gravitational instability in, for example, \citet{Boley:2007p642}.

Also, the stability or fragmentation of circumstellar disks, which leads to the formation of planets, brown dwarfs, or binary stars, depends mainly on two temperature-dependent criteria,
the Toomre criterion \citep{Toomre:1964p677} and
the cooling timescale \citep{Gammie:2001p25}.
Based on these criteria, semi-analytical studies of massive accretion disks \citep{Kratter:2006p428, Kratter:2008p196, Vaidya:2009p436} derive the disk's stability and its fragmentation properties.
The study by \citet{Vaidya:2009p436} includes gas and dust opacities to obtain dynamical quantities in the midplane of disks around typical massive stars and concludes that viscous heating is the dominant phenomenon in destroying dust up to roughly 10~AU from the central star. 
This location of the dust evaporation front is in good agreement with the dynamical disk-formation studies by 
\citet{Kuiper:2013p17358},
which also take the effect of optically thick gas (disks) around massive stars into account.

Planetary companions may form in the accretion disk directly by disk fragmentation or 
by core-accretion \citep{Pollack:1996p17324, Alibert:2004p17254, Alibert:2005p17281}.
Based on the core-accretion scenario, statistical properties of extrasolar planets can be determined using an extrasolar planet population synthesis model \citep{Mordasini:2009p17238, Mordasini:2009p17253, Alibert:2011p17315, Mordasini:2012p17258}.
Effects of the disk's temperature distribution (due to irradiation) on the outcome of such planet population synthesis models are studied in \citet{Fouchet:2012p17303}.

After a companion has formed in an accretion disk, its radial drift or migration plays a major role in the further evolution of the solid body-disk interacting system.
In this late stage of disk evolution, the effect of radiation on migration is an ongoing problem in modern astrophysical research \citep[see, e.g.,][]{Bitsch:2012p16070}.

Summing up,
a reliable determination of the disk's temperature distribution is a crucial ingredient of current and future (magneto-) hydrodynamical studies of all evolutionary epochs of circumstellar disks.
To account for the proper temperature distribution, (magneto-) hydrodynamics studies of irradiated circumstellar disks require an accurate treatment of radiation transport. 

At the same time this accurate radiation transport should desirably be as fast as the (magneto-) hydrodynamics to allow for an efficient usage of computing resources. Such fast radiation transport methods imply the acceptance of far-reaching approximations, e.g.~neglecting frequency dependence or making use of the diffusion approximation.
Although \citet{Harries:2011p1968} recently combined Monte-Carlo radiation transport with hydrodynamics, they conclude that the Monte-Carlo calculations are so expensive in cpu time that the time spent for the hydrodynamics part of the simulations becomes negligible.
By consuming roughly the same computing resources for the radiation transport than for the gas motion itself, approximate radiation transport methods make it possible, for example, to study the system under investigation for much longer evolution in time, run a whole set of simulations to scan the parameter space, or achieve higher resolution.

Here, we extend our previous comparison test of radiation transport methods \citep{Kuiper:2010p586} by considering three types of approximate radiation transport methods and expanding the setup of the disks to very high optical depth.
As physical test setup, we choose the standard radiation benchmark tests by \citet{Pascucci:2004p327} and \citet{Pinte:2009p418}.
We perform simulations for a wide range of optical depths of the disk's midplane from $\tau_\mathrm{550nm}=0.1$ up to $\tau_\mathrm{810nm}=1.22\times10^{+6}$.
For comparison, we use results from the 
Monte-Carlo code RADMC
\citep{Dullemond:2000p686, Dullemond:2004p523}.

We determine the accuracy 
both of the gray and frequency-dependent hybrid radiation transport method, introduced in \citet{Kuiper:2010p586},
and of
the gray flux-limited diffusion (FLD) approximation, which denotes a standard technique in current astrophysical research.
In the hybrid method, stellar irradiation is computed in a ray-tracing (RT) step and the thermal (re-)emission of dust is shifted to a gray FLD solver.
Previous studies using the hybrid radiation transport approach covered 
the formation of massive stars and their radiation pressure feedback \citep{Kuiper:2010p541, Kuiper:2011p349, Kuiper:2013p17358},
along with
the stability of radiation-pressure-dominated outflow cavities around forming high-mass stars \citep{Kuiper:2012p1151}.
The latter study emphasizes the importance of the long-range effect of stellar irradiation to compute the stability of the outflow cavities.
It shows that the usage of the gray FLD approximation yields an underestimate of the radiative force in the expanding cavity shell by 1-2 orders of magnitude (depending on the size of the cavity and the stellar parameters). 
These results were later confirmed by radiation-hydrodynamics simulations of \citet{Harries:2012p15058} and radiation transport models of \citet{Owen:2012p16415}.
Another example of the importance of such radiative long-range effects was already given in three-dimensional radiation-hydrodynamics studies of proto-planetary disks by \citet{Boley:2007p642}.

For an historical overview of the development of approximate radiation transport methods we refer the reader to the introduction of \citet{Kuiper:2010p586}.

\section{Methods}
\label{sect:method}
The equilibrium temperature for static disk configurations is computed using different kinds of radiation transport methods, including fast approximate radiation transport solvers and a highly accurate Monte-Carlo code.

As approximate radiation transport solvers, we use 
the gray FLD approximation,
the gray irradiation + gray FLD approximation, and
the frequency-dependent irradiation + gray FLD approximation.
We have described the details of these techniques and their actual numerical implementation in our radiation transport module called MAKEMAKE in 
\citet{Kuiper:2010p586} and
\citet{Kuiper:2012p1151}.
Due to their speed, these approximate radiation transport solvers are ideally suited to be used in magneto-hydrodynamics simulations as well.

As a comparison code, we use 
the Monte-Carlo based radiation transfer code RADMC, which is described in
\citet{Dullemond:2000p686} and \citet{Dullemond:2004p523}.
The general solver method of RADMC is based on \citet{Bjorkman:2001p645}.
The code is also described briefly in the original radiation benchmark tests of
\citet{Pascucci:2004p327} and \citet{Pinte:2009p418}.
While scattering can be handled by this code, it is simply switched off because it is also neglected in the approximate radiation transport methods for magneto-hydrodynamics simulations. 
For the given disk and envelope configurations of moderate optical depth $\tau_\mathrm{550nm}=100$, scattering would increase the temperature in the irradiated parts by about 2\% in the optically thin envelope and up to a maximum of 19\% in the optically thick inner rim of the disk's midplane (up to an optical depth of about unity) due to higher extinction. 
The more effectively shielded outer regions of the disk would be about 10\% cooler.

\section{Physical setup, numerical configuration, and runs performed}
\begin{figure}[tb]
\begin{center}
\fbox{
\includegraphics[width=0.45\textwidth]{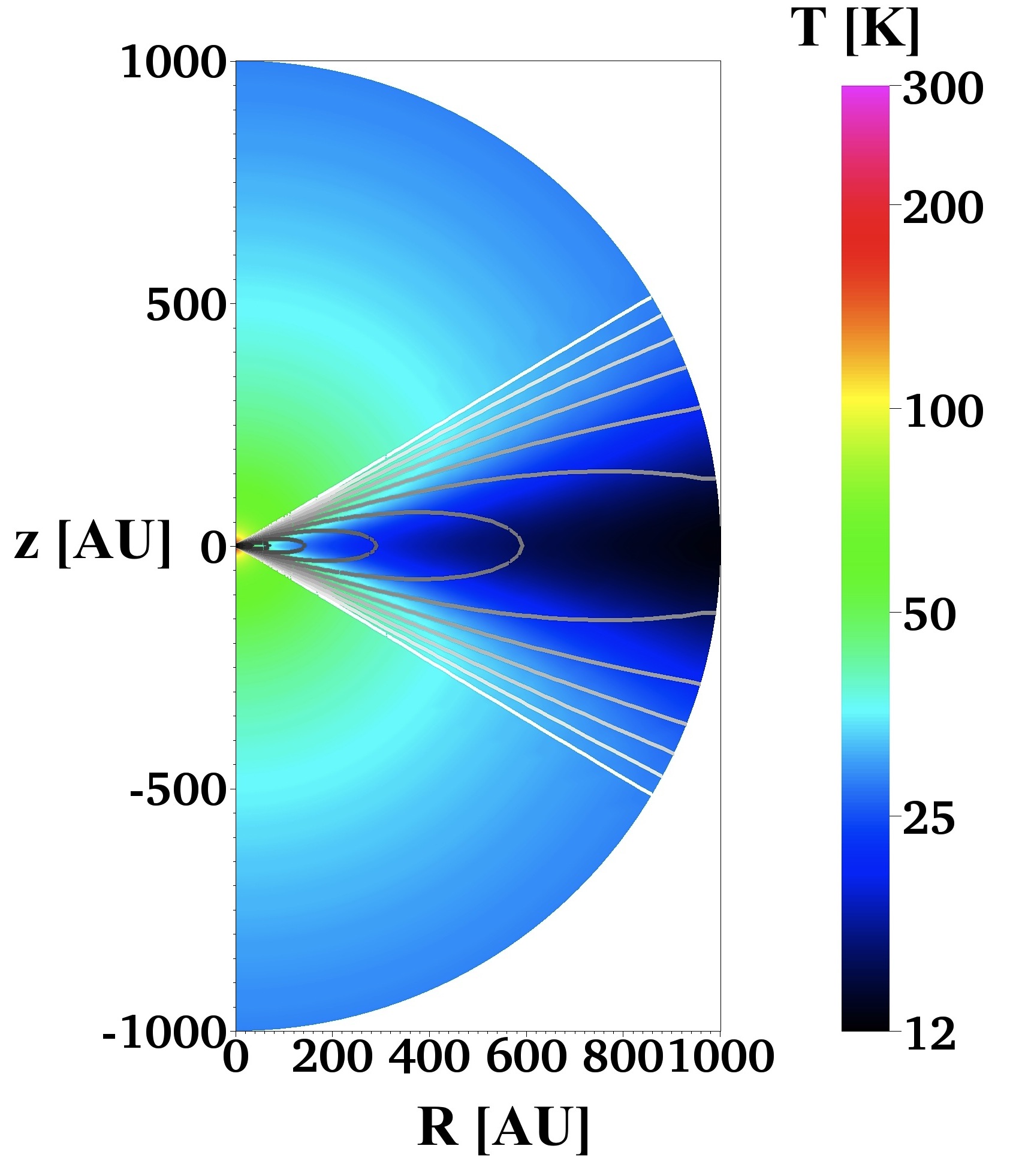}
}
\caption{
Visualization of the disk configuration and final temperature for the case $\tau_\mathrm{550nm} = 100$.
Solid lines denote iso-density contours.
Colors denote the final temperature distribution.
}
\label{fig:VisIt}
\end{center}
\end{figure}
\subsection{Physical setup}
The setup is adopted from the original standard radiation benchmark tests \citet{Pascucci:2004p327} and \citet{Pinte:2009p418}.
It describes a slightly flared circumstellar disk within an optically thin envelope around a central proto-star. 
Such a flaring is assumed to be a natural consequence of the heating \citep[e.g.,][]{Kenyon:1987p17217, Chiang:1997p17115}, 
but generally geometries of circumstellar disks can have a large variety \citep{Dullemond:2004p1401}.
The setup is axially symmetric.
Hence, the radiation transport is done in 2D.
The gas density distribution of the circumstellar disk is given in cylindrical coordinates $(r,z)$ by
\begin{equation}
\rho(r,z) = \rho_0~f_1(r)~f_2(r,z),
\end{equation}
with the radially and vertically dependent functions
\begin{equation}
f_1(r) = r_\mathrm{d} / r \mbox{\hspace{2mm} and  \hspace{2mm}} f_2(r,z) = \exp\left(-\frac{\pi}{4} \left(\frac{z}{h(r)}\right)^2\right)
\end{equation}
making use of the abbreviations 
$h(r) = z_\mathrm{d} (r/r_\mathrm{d})^{1.125}$,
$r_\mathrm{d} = 500$~AU, and
$z_\mathrm{d} = 125$~AU.
The normalization $\rho_0$ of the density setup is chosen to define different optical depths $\tau$ through the midplane of the corresponding circumstellar disk.
We compute problems scanning seven orders of magnitude in optical depth.

The central star is assumed to radiate as a black body.
In the \citet{Pascucci:2004p327} radiation benchmark test ($\tau_\mathrm{550nm} \le 100$),
the star has a radius of $R_* = 1 \mbox{ R}_\odot$ and
a temperature of $T_* = 5800$~K.
In the \citet{Pinte:2009p418} radiation benchmark test ($\tau_\mathrm{810nm} \ge 1000$),
the star has a radius of $R_* = 2 \mbox{ R}_\odot$ and
a temperature of $T_* = 4000$~K.
A visualization of the disk setup and the final temperature distribution is given in Fig.~\ref{fig:VisIt}.

\begin{figure}[htbp]
\begin{center}
\includegraphics[width=0.5\textwidth]{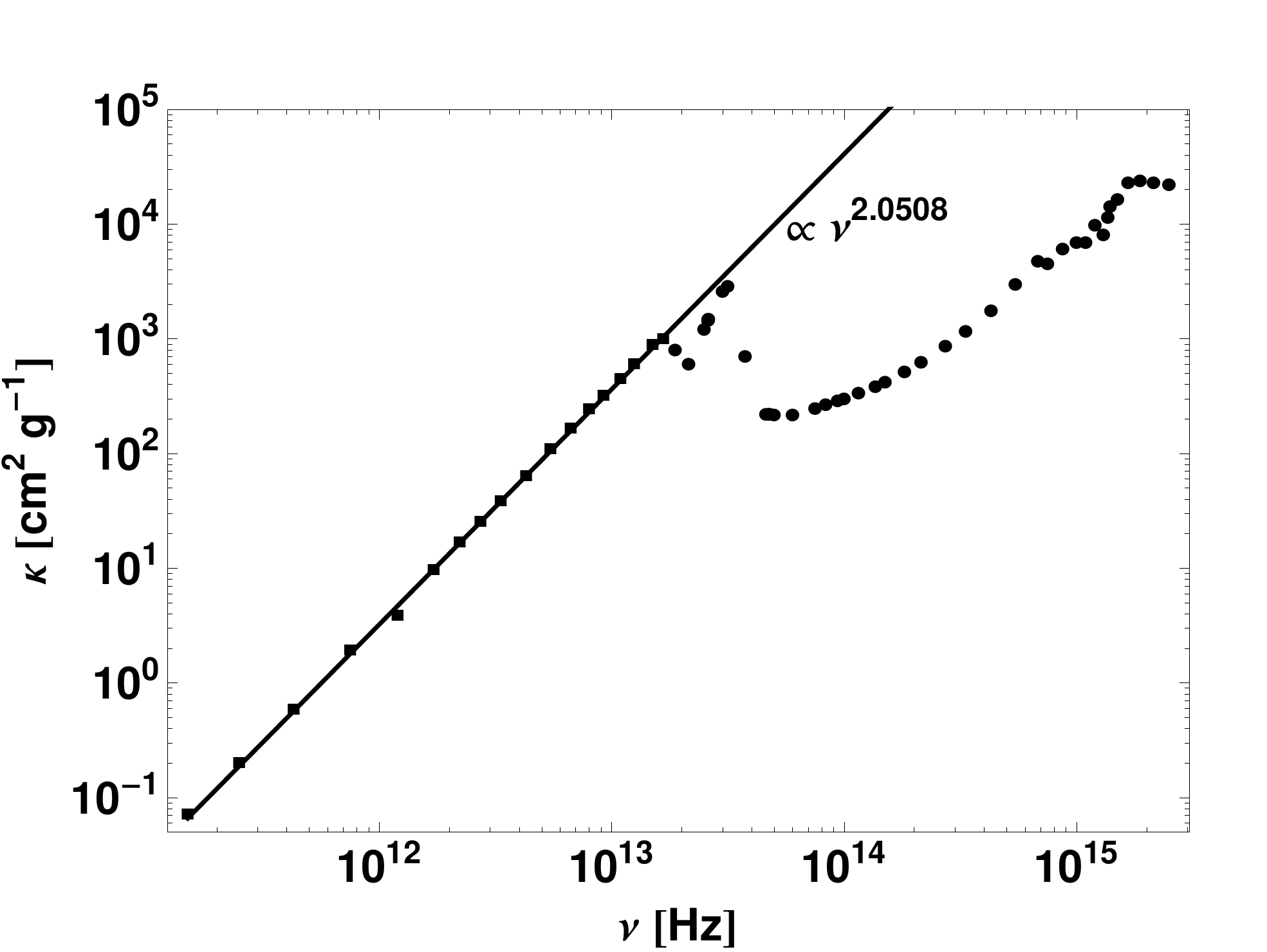}
\caption{
Frequency-dependent opacities from \citet{Draine:1984p118}.
The solid line denotes an exponential fit to the long wavelength regime.
}
\label{fig:opacity}
\end{center}
\end{figure}

The opacity table used is the same as in the original benchmark test of \citet{Pascucci:2004p327} taken from \citet{Draine:1984p118}.
These opacities are shown in Fig.~\ref{fig:opacity}.
The dust-to-gas mass ratio is set constant to 1\%.
Scattering is neglected.

\subsection{Numerical configuration}
\label{sect:configuration}
In the \citet{Pascucci:2004p327} radiation benchmark test ($\tau_\mathrm{550nm} \le 100$),
the computational domain goes from $R_\mathrm{min} = 1$~AU up to $R_\mathrm{max} = 1000$~AU.
In the \citet{Pinte:2009p418} radiation benchmark test ($\tau_\mathrm{810nm} > 1000$),
the computational domain goes from 0.1~AU up to 400~AU.
In both setups, the polar angle covers the full domain from 0 to $180$ degrees.

The boundary conditions in the polar direction are chosen so as not to allow any radiative flux over the polar axis.
The boundary condition at the outer radial direction is chosen to allow for a radiative flux out of the computational domain computed in the optically thin limit.
The boundary condition for the thermal radiation field at the inner radial direction is in the RT simulations given by a low value of the thermal radiation field (the total radiation field at the inner disk rim is dominated by the irradiation).
The boundary condition for the thermal radiation field at the inner radial direction is in the simulations using only FLD determined from the surface temperature of the central star assuming an optically thin limit between the stellar surface and the inner disk rim (as in the original benchmark tests).

The grids in spherical coordinates are chosen to resolve the local optical depths of the circumstellar disks in the polar direction in order to guarantee the computation of correct cooling properties.
For all cases up to $\tau_\mathrm{810nm}=1.22\times10^{+3}$, 
the grid consists of $n_r \times n_\theta = 128 \times 360$ cells.
For $\tau_\mathrm{810nm}=1.22\times10^{+4}$, 
the grid has $n_r \times n_\theta = 64 \times 2880$ grid cells.
For $\tau_\mathrm{810nm}=1.22\times10^{+6}$, 
the grid has $n_r \times n_\theta = 16 \times 10^5$ grid cells.

We note that results of the same accuracy are obtained by also using $128\times360$ grid cells for the higher optical depths runs.
However, they make use of a numerical `trick' similar to the one described in \citet{Pinte:2009p418} for the ProDiMo code, where the density distribution was altered to limit the optical depth.
Here, we promote a modified algorithm of the same idea:
The resolution of a $128\times360$ grid is sufficient also for high optical depth runs, if an upper limit of the optical depth of $\tau_\mathrm{max} \le 1$ is applied to each individual grid cell during the radiation transport.
Although we focus in the following discussion on the results of the high resolution grids only, a solution of a simulation run using this numerical trick is shown for comparison for the most optically thick case of $\tau_\mathrm{810nm} = 1.22\times10^6$ in Fig.~\ref{fig:tau1e+6}.

The numerical configuration of the
Monte-Carlo code RADMC
is given in
\citet{Pascucci:2004p327} and \citet{Pinte:2009p418} with the difference that we do not consider scattering here.

\begin{table*}[tb]
\begin{center}
\begin{tabular}{l l c c c}
Run label & Comparison & $\tau$ & Radiation Transport & $(\Delta T)_\mathrm{max}$~[\%]\\
\hline
Pa-1e-1-grayRT+FLD & \multirow{4}{20mm}{Pascucci et al.} & \multirow{4}{5mm}{$10^{-1}$} & gray RT + FLD & 3 \\
Pa-1e-1-freqRT+FLD & & & freq. RT + FLD & 3 \\
Pa-1e-1-FLD & & & FLD & 55 \\
Pa-1e-1-MC & & & Monte-Carlo & 1 \\
\hline
Pa-1e+2-grayRT+FLD & \multirow{4}{20mm}{Pascucci et al.} & \multirow{4}{5mm}{$10^{+2}$} & gray RT + FLD & 54 \\
Pa-1e+2-freqRT+FLD & & & freq. RT + FLD & 16 \\
Pa-1e+2-FLD & &  & FLD & 42 \\
Pa-1e+2-MC & &  & Monte-Carlo & 14 \\
\hline
Pi-1e+3-grayRT+FLD & \multirow{4}{20mm}{Pinte et al.} & \multirow{4}{5mm}{$10^{+3}$} & gray RT + FLD & 48 \\
Pi-1e+3-freqRT+FLD & & & freq. RT + FLD & 48 \\
Pi-1e+3-FLD & &  & FLD & 282 \\
Pi-1e+3-MC & &  & Monte-Carlo & 19 \\
\hline
Pi-1e+4-grayRT+FLD & \multirow{4}{20mm}{Pinte et al.} & \multirow{4}{5mm}{$10^{+4}$} & gray RT + FLD & 46 \\
Pi-1e+4-freqRT+FLD & & & freq. RT + FLD & 46 \\
Pi-1e+4-FLD & &  & FLD & 284 \\
Pi-1e+4-MC & &  & Monte-Carlo & - \\
\hline
Pi-1e+6-grayRT+FLD & \multirow{4}{20mm}{Pinte et al.} & \multirow{4}{5mm}{$10^{+6}$} & gray RT + FLD & 37 \\
Pi-1e+6-freqRT+FLD & & & freq. RT + FLD & 37 \\
Pi-1e+6-FLD & &  & FLD & 289 \\
Pi-1e+6-MC & &  & Monte-Carlo & 42 \\
\hline
\end{tabular}
\end{center}
\caption{
Overview of simulations performed.
The columns denote
the run label,
the reference to the original benchmark test (either \citet{Pascucci:2004p327} or \citet{Pinte:2009p418}),
the optical depth of the disk's midplane (either $\tau_\mathrm{550nm}$ or $\tau_\mathrm{810nm}$, respectively),
the radiation transport method used in the simulation, and
the resulting temperature deviation in percent.
The deviations are computed with respect to the Monte-Carlo solution by RADMC.
The temperature deviations given for the Monte-Carlo runs are taken from \citet{Pascucci:2004p327}, Figs.~4 and 5, or \citet{Pinte:2009p418}, Fig.~10, with respect to the deviation to the other radiation transport codes participating in the original benchmark tests.
A visualization of the final maximum temperature deviations as a function of optical depth is given in Fig.~\ref{fig:deltaT}.
}
\label{tab:run-table}
\end{table*}

\subsection{Runs performed}
We computed the equilibrium temperature distribution in the disk and envelope using different radiation transport methods.
We studied
the gray FLD approximation,
the gray irradiation + gray FLD approximation, and
the frequency-dependent irradiation + gray FLD approximation.
As a reference solution, we used results obtained with the Monte-Carlo code RADMC.
We performed simulations for various optical depths of the disk's midplane from $\tau_\mathrm{550nm}=0.1$ up to $\tau_\mathrm{810nm}=1.22\times10^{+6}$.
An overview of runs performed is given in Table~\ref{tab:run-table}.

\section{Simulation results}
\begin{figure}[htbp]
\begin{center}
\includegraphics[width=0.5\textwidth]{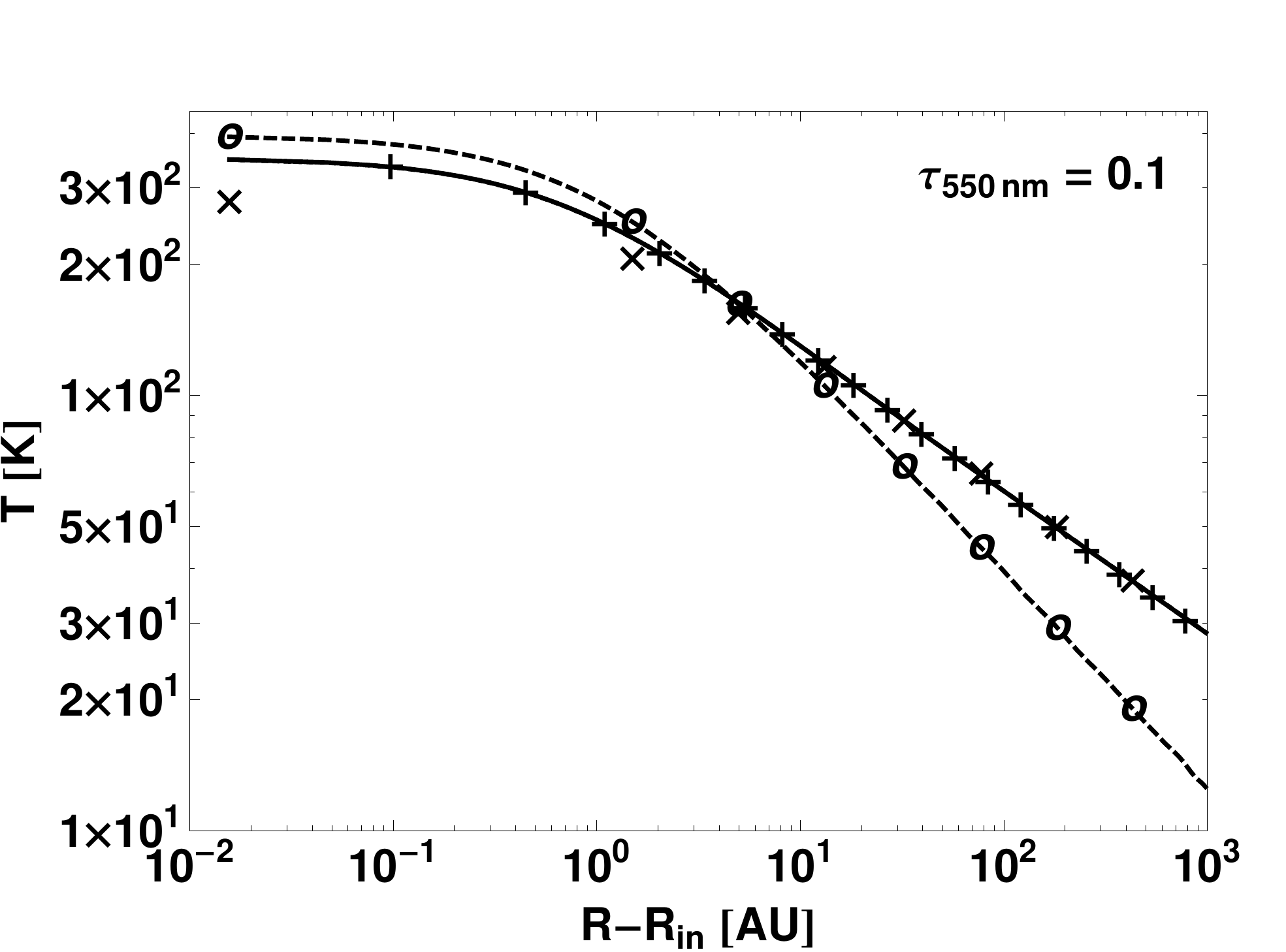}
\includegraphics[width=0.5\textwidth]{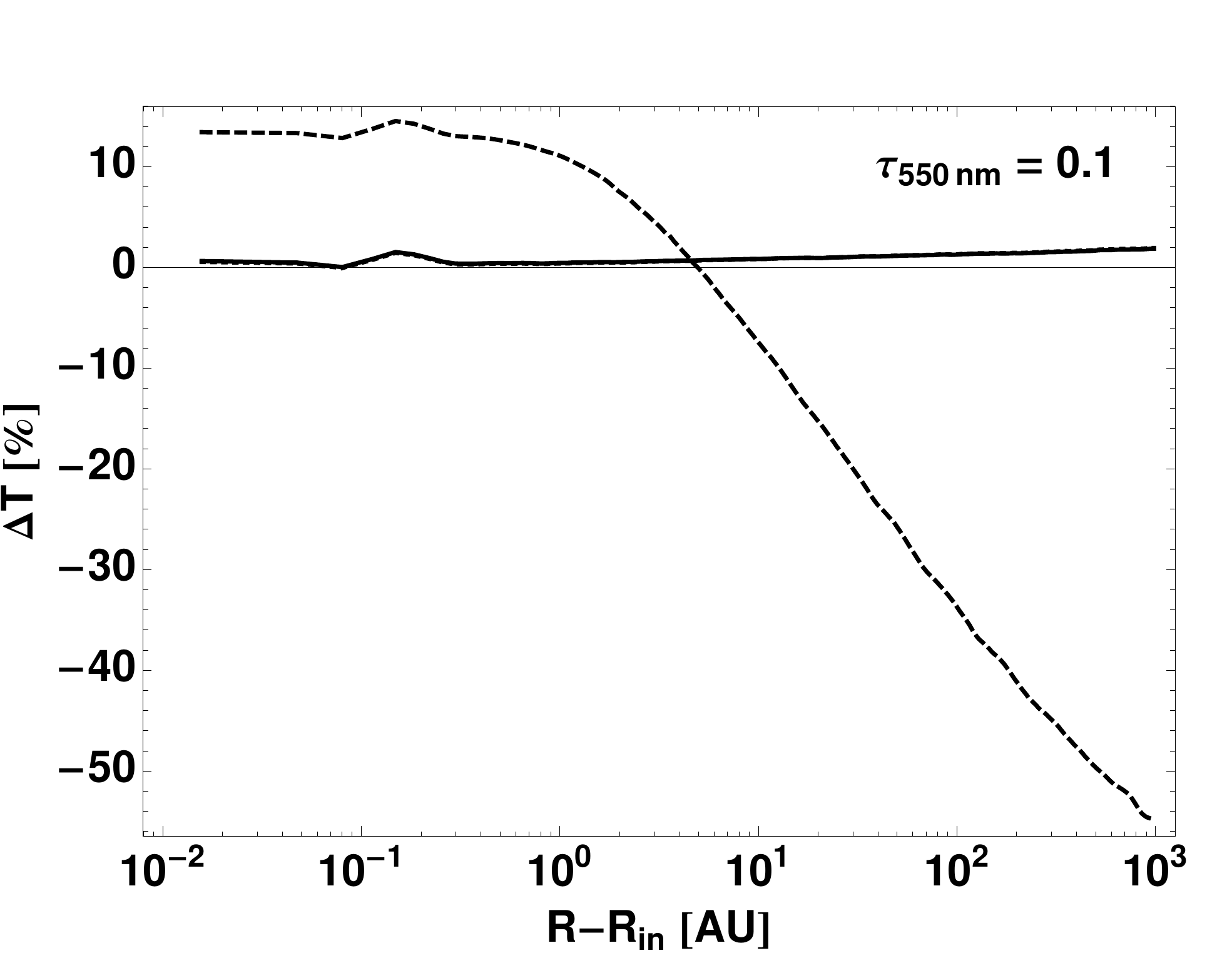}
\caption{
Temperature profiles (upper panel) in the midplane of the circumstellar disk
for the case of low optical depth $\tau_{550\mathrm{nm}}=0.1$.
In the lower panel the deviations of the three different radiation transport methods to the comparison code result are displayed.
Solid lines denote results for the frequency-dependent RT + gray FLD method.
Dotted lines denote results for the gray RT + gray FLD method.
Dashed lines denote results for the gray FLD method.
Pluses ``+'' denote results for the comparison code RADMC.
In this optical thin case, the results for gray RT + gray FLD are almost identical to the frequency-dependent RT + gray FLD.
Here, analytical estimates for the optically thin approximation are given: 
crosses ``x'' mark the analytical estimate by \citet{Spitzer:1978p776} for irradiated regions far away from the star,
circles ``o'' mark the analytical estimate in the gray and isotropic approximation ($T \propto r^{-1/2}$).
}
\label{fig:tau1e-1}
\end{center}
\end{figure}

\begin{figure}[htbp]
\begin{center}
\includegraphics[width=0.5\textwidth]{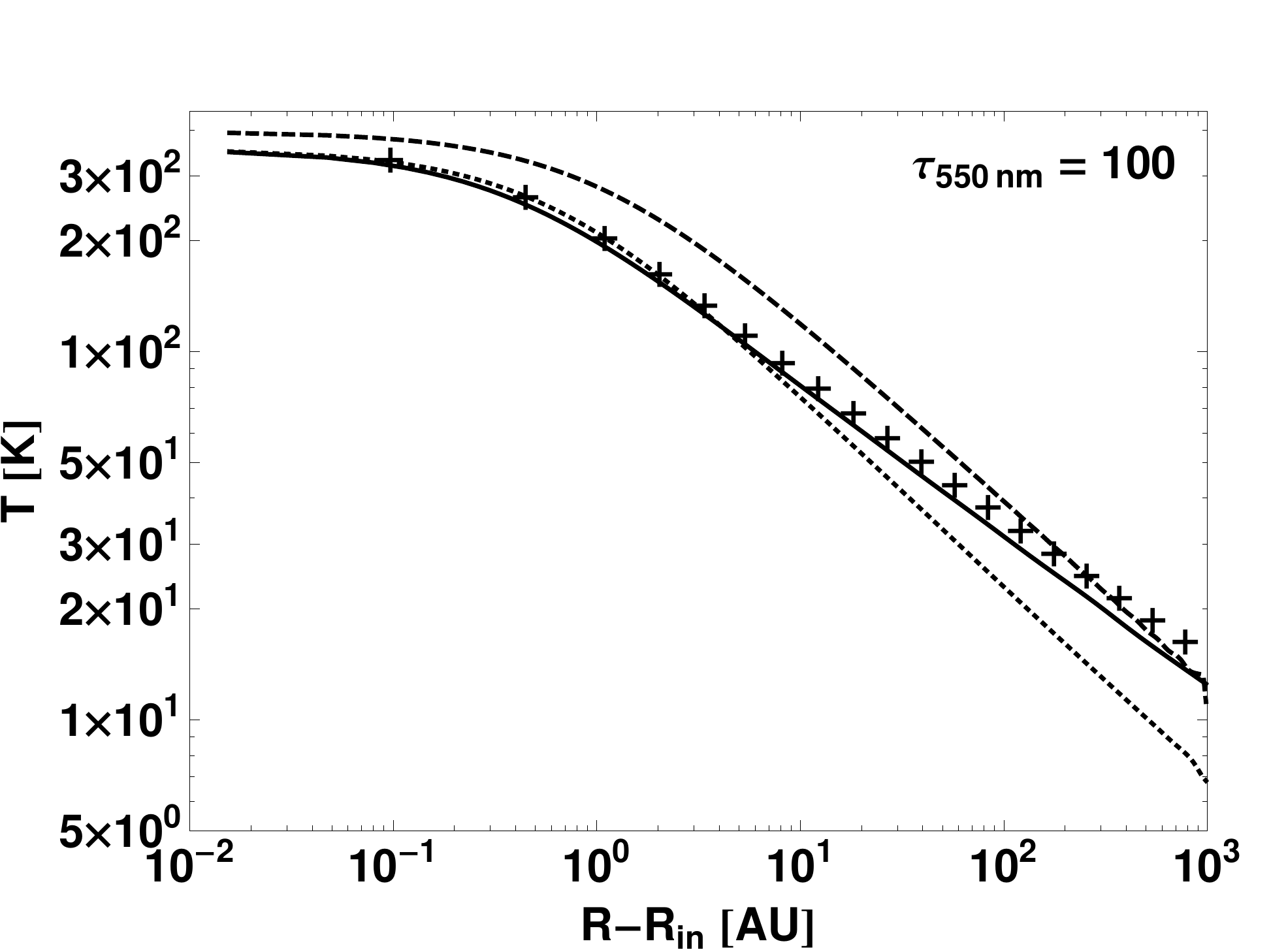}
\includegraphics[width=0.5\textwidth]{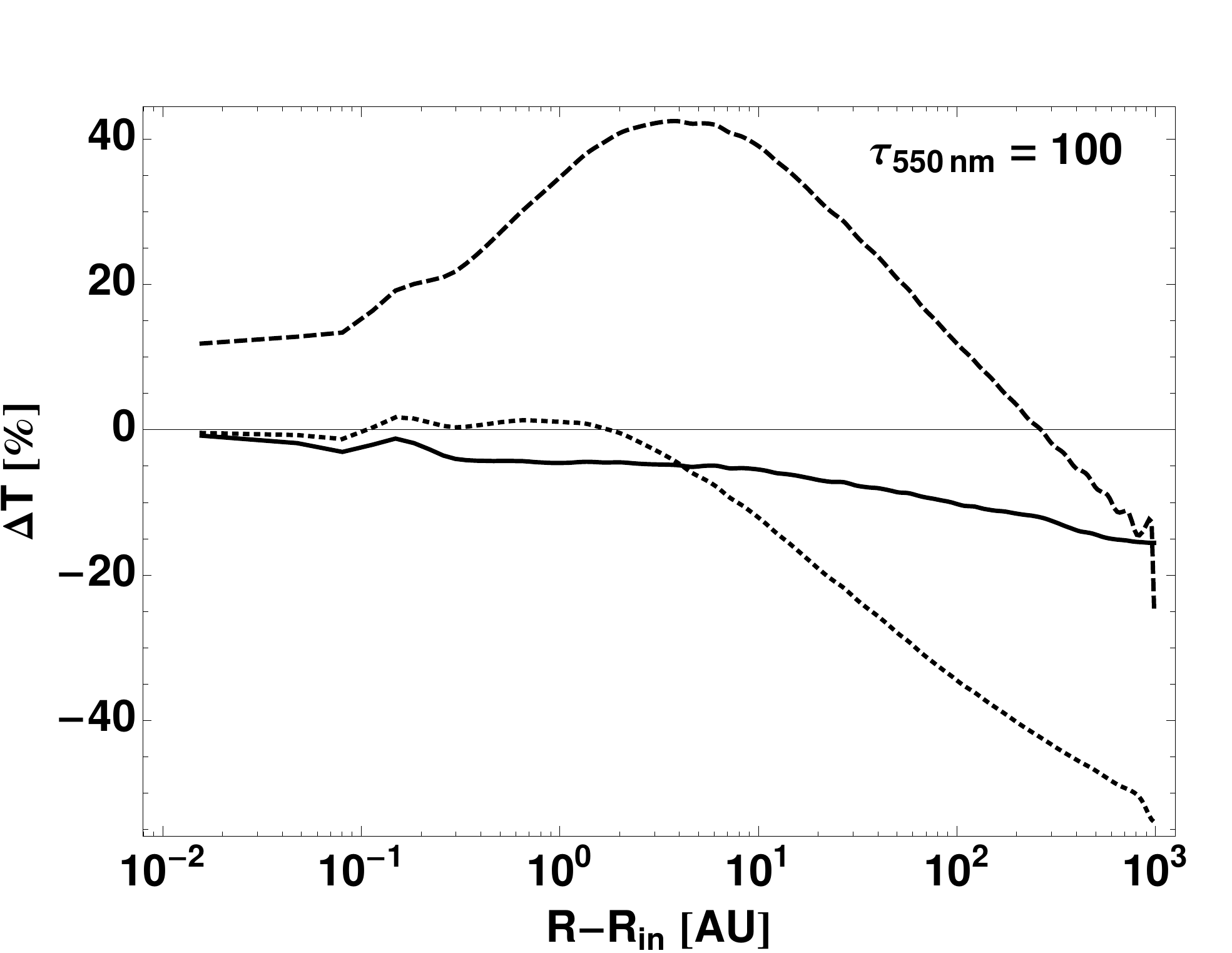}
\caption{
Same as Fig.~\ref{fig:tau1e-1} for simulation runs with $\tau_{550\mathrm{nm}}=10^2$.
}
\label{fig:tau1e+2}
\end{center}
\end{figure}

\begin{figure}[htbp]
\begin{center}
\includegraphics[width=0.5\textwidth]{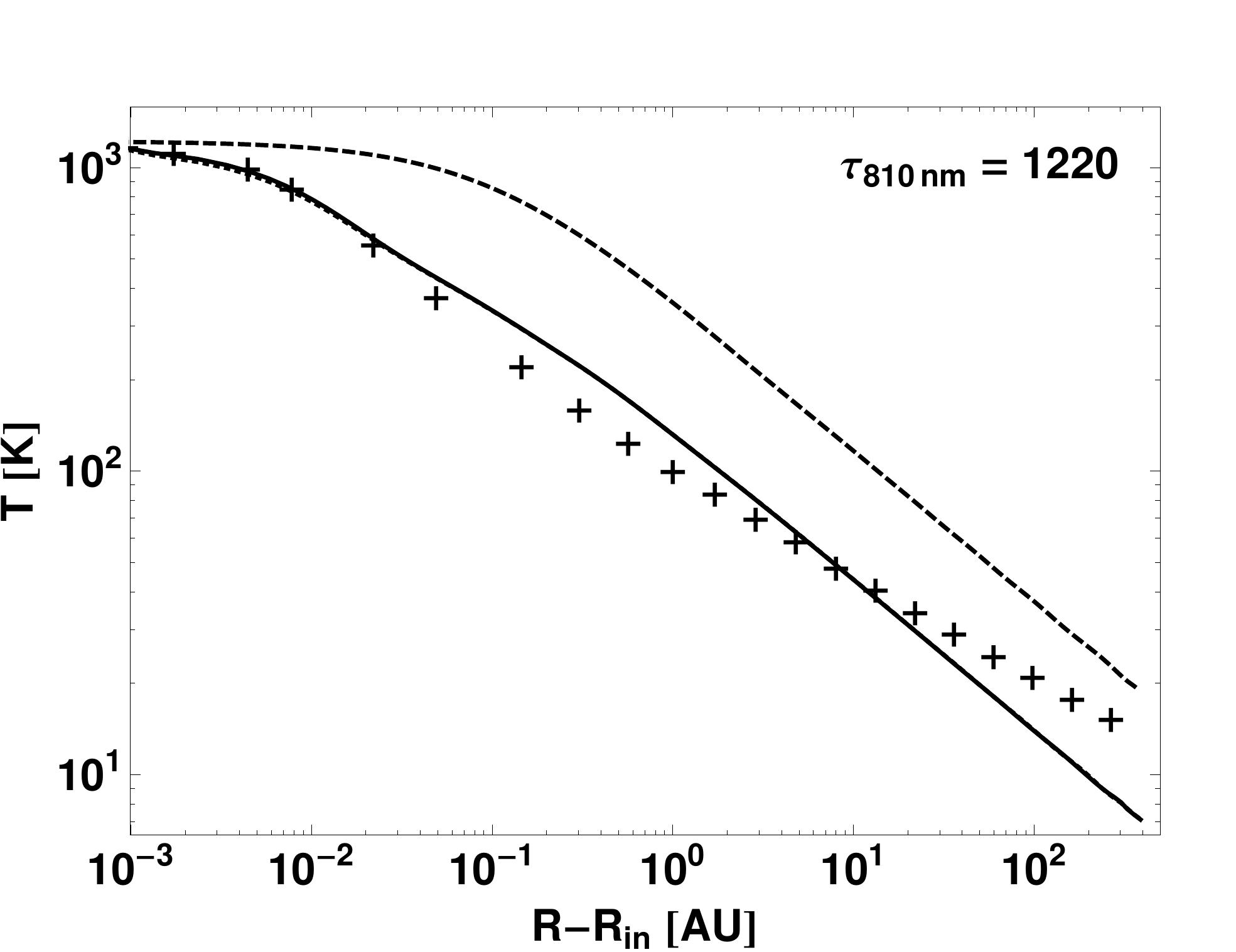}
\includegraphics[width=0.5\textwidth]{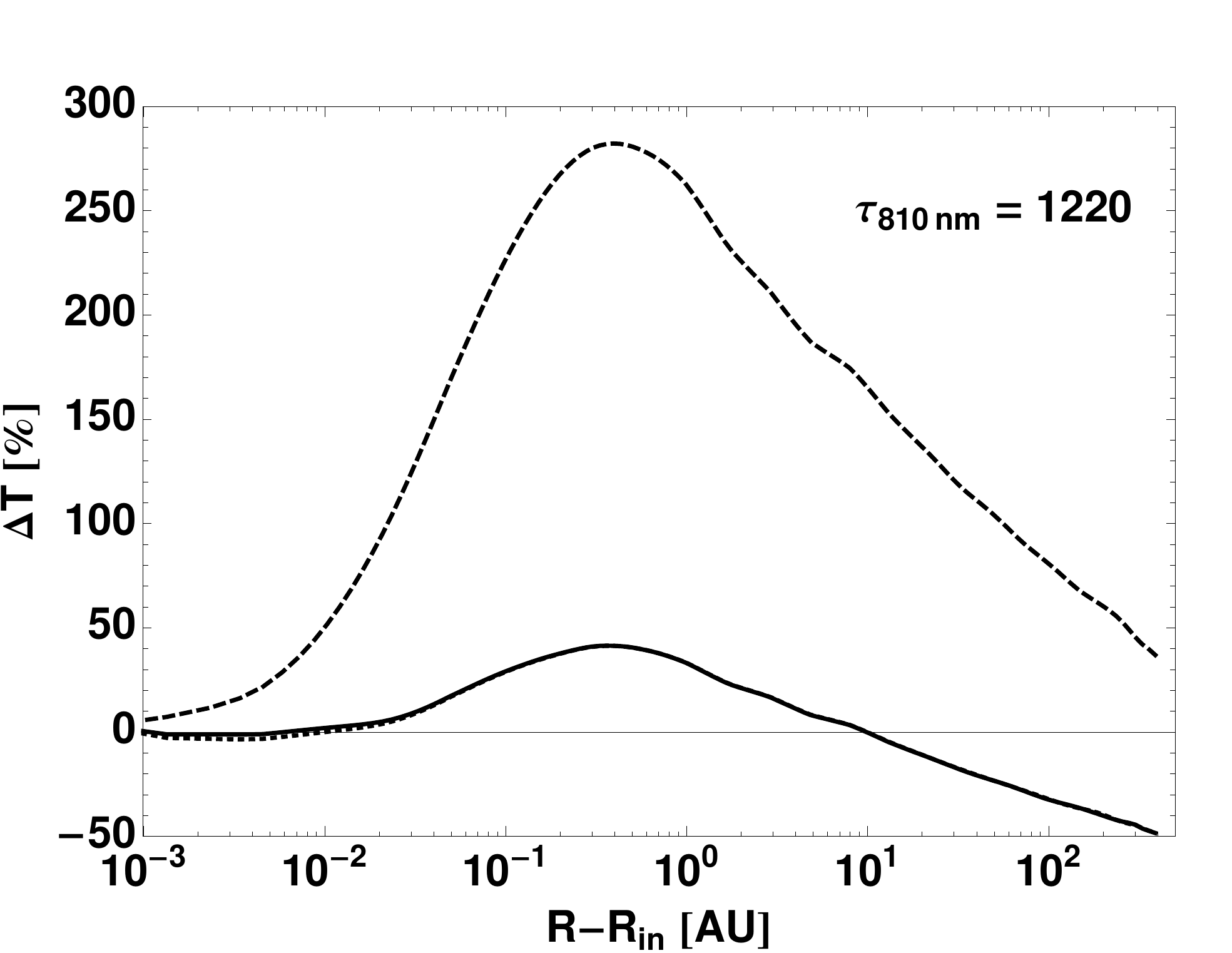}
\caption{
Same as Fig.~\ref{fig:tau1e-1} for simulation runs with $\tau_{810\mathrm{nm}}=1.22\times10^3$.
In this fairly optical thick case, the results for gray RT + gray FLD are almost identical to the frequency-dependent RT + gray FLD, small differences are visible at the innermost rim of the disk only.
}
\label{fig:tau1e+3}
\end{center}
\end{figure}

\begin{figure}[htbp]
\begin{center}
\includegraphics[width=0.5\textwidth]{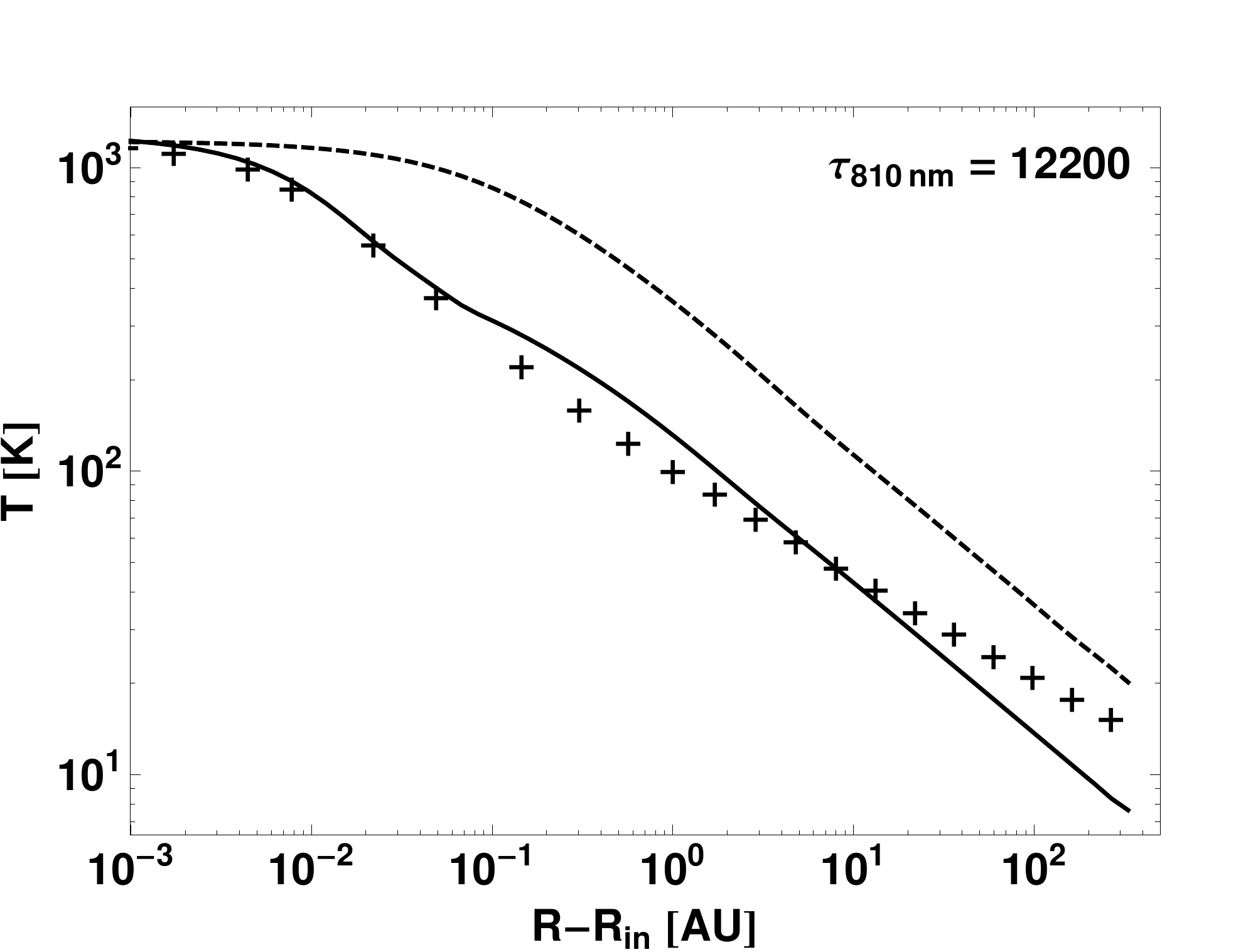}
\includegraphics[width=0.5\textwidth]{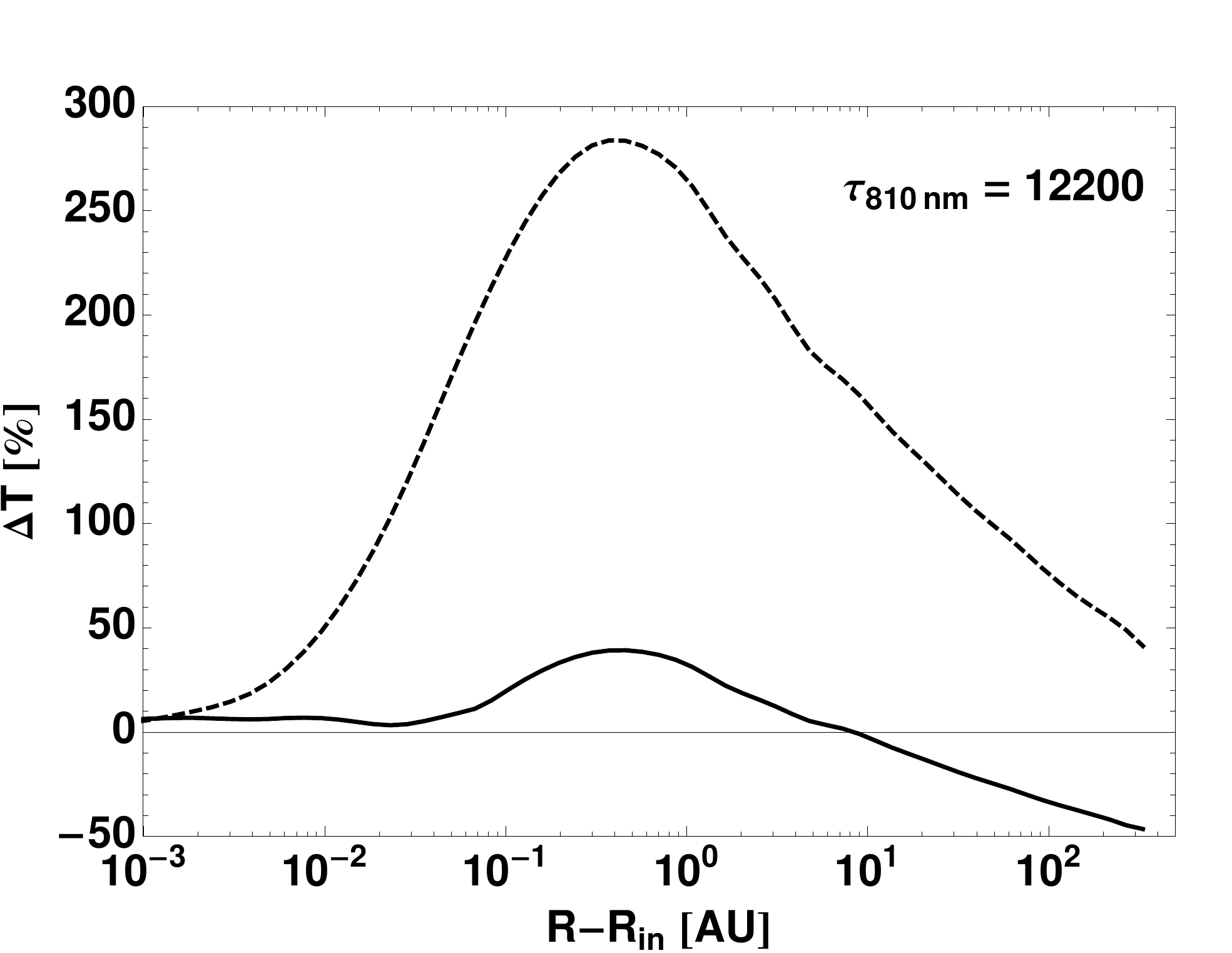}
\caption{
Same as Fig.~\ref{fig:tau1e-1} for simulation runs with $\tau_{810\mathrm{nm}}=1.22\times10^4$.
In this highly optical thick case, the results for gray RT + gray FLD are identical to the frequency-dependent RT + gray FLD.
}
\label{fig:tau1e+4}
\end{center}
\end{figure}

\begin{figure}[htbp]
\begin{center}
\includegraphics[width=0.5\textwidth]{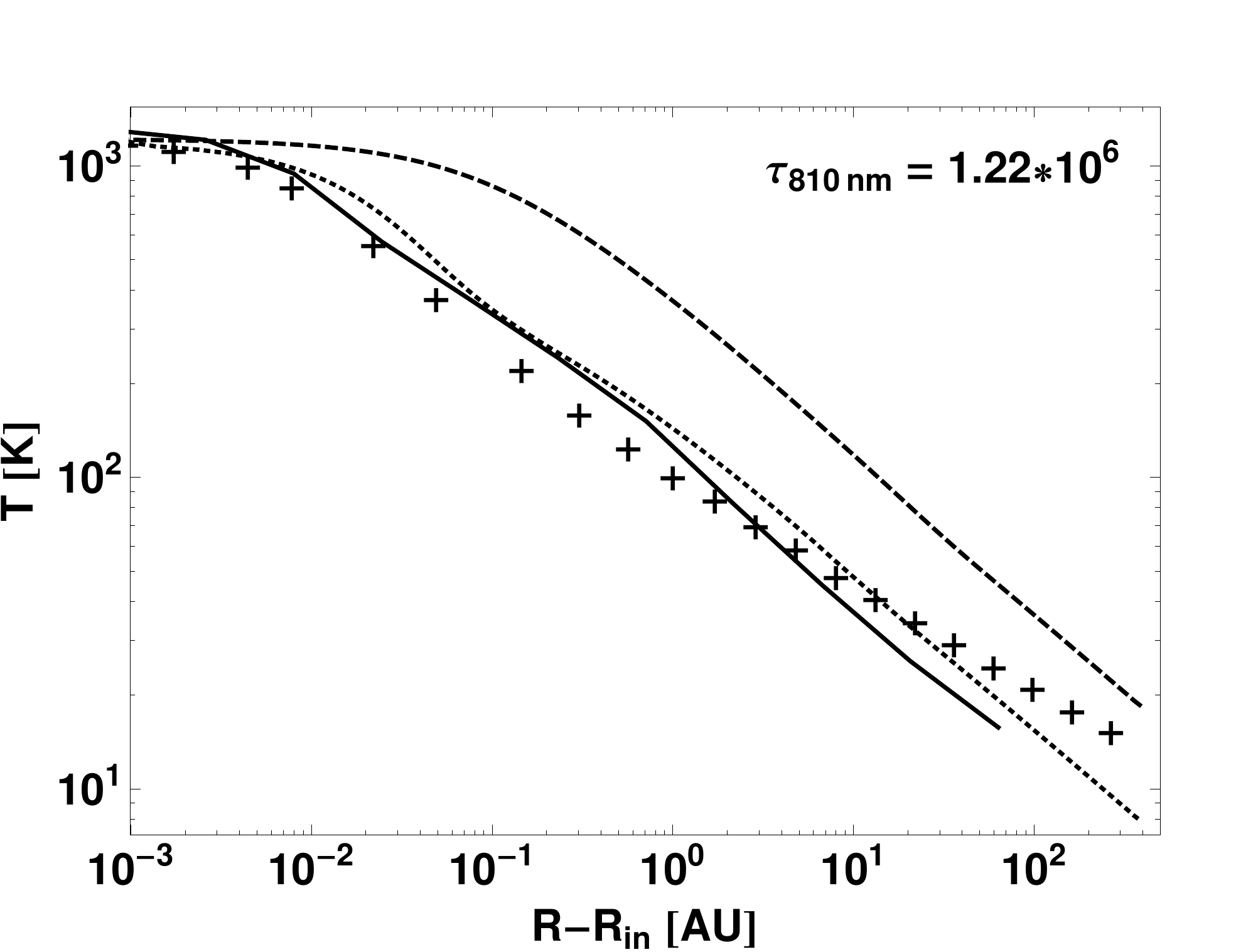}
\includegraphics[width=0.5\textwidth]{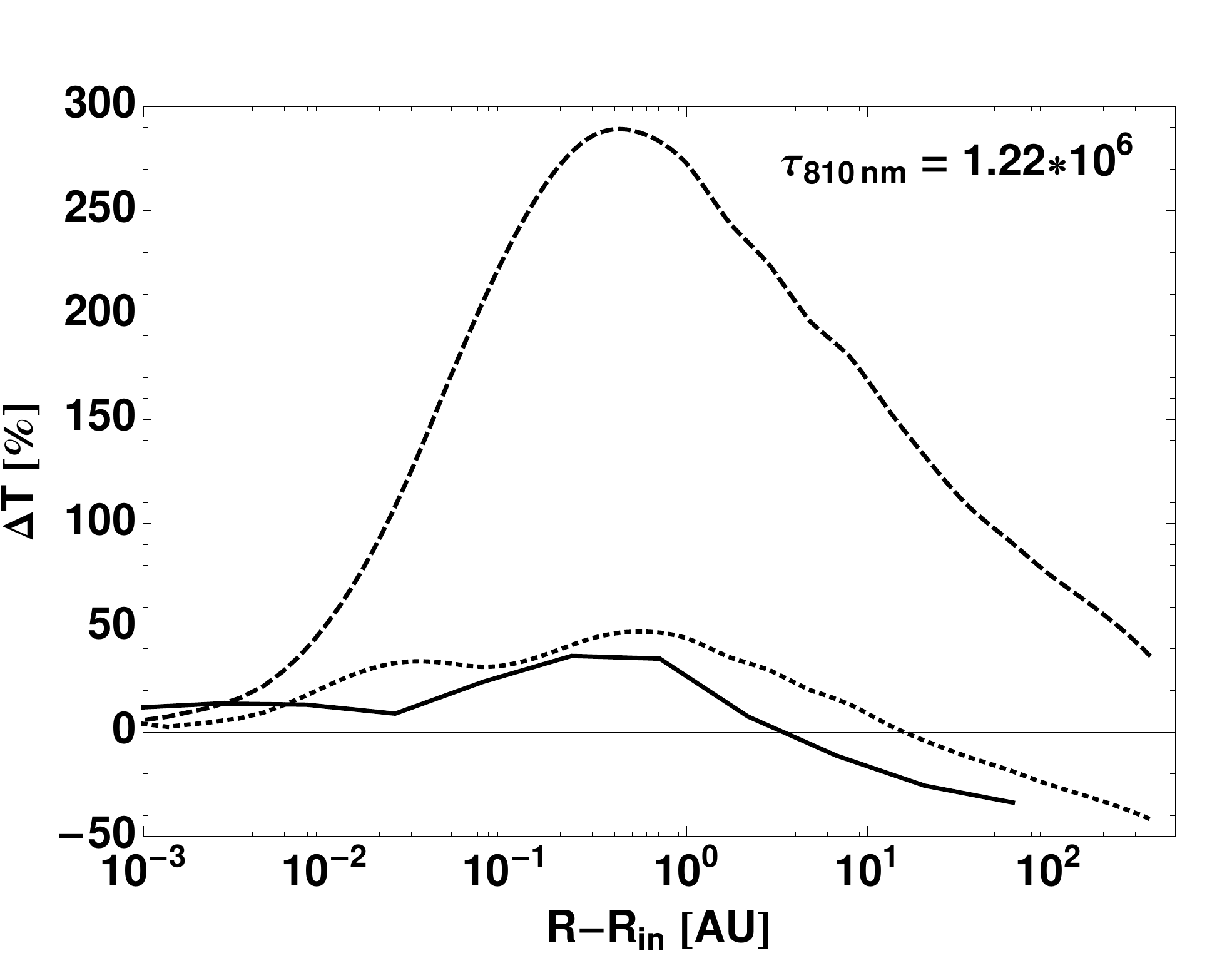}
\caption{
Same as Fig.~\ref{fig:tau1e-1} for simulation runs with $\tau_{810\mathrm{nm}}=1.22\times10^6$.
In this highly optical thick case, the results for gray RT + gray FLD are identical to the frequency-dependent RT + gray FLD.
Here,
the dotted lines represent the results for the frequency-dependent RT + gray FLD method in a low-resolution simulation if an upper limit of the optical depth per grid cell of $\tau_\mathrm{max} \le 1$ is used (see Sect.~\ref{sect:configuration} for a description of this numerical trick).
}
\label{fig:tau1e+6}
\end{center}
\end{figure}

We compared the resulting temperature profiles through the disk's midplane for the different radiation transport methods depending on the optical depth of the circumstellar disk.
The final accuracy of the different approximate radiation transport methods was computed as the deviation to the comparison code results by RADMC.
These deviations and the standard deviation of the comparison code with respect to the other radiation transport codes participating in the original benchmark tests of  \citet{Pascucci:2004p327} and \citet{Pinte:2009p418} are given in the last column of Table~\ref{tab:run-table}.
Visualizations of the individual temperature profiles and the deviations to the reference code are given in Figs.~\ref{fig:tau1e-1} to \ref{fig:tau1e+6}.

\section{Discussion}
In the following, we discuss the simulation results and their implications, depending on the different regimes of optical depth.
Afterwards, we discuss the validity of the various approximations used in the different radiation transport methods described above.

\subsection{Different regimes of optical depth}
We begin our discussion with the two limiting cases of very optically thin and thick media.
The regime of moderate optical depth is addressed afterwards.

\subsubsection{The most optically thin regime}
In the optically thin limit, the temperature profile can be accurately estimated analytically.
As given in \citet{Spitzer:1978p776}, the local dust temperature in irradiated regions far away from the central star ($r \gg R_*$) is presented by 
\begin{equation}
\label{eq:Spitzer}
T(r) = \left(\frac{R_\mathrm{min}}{2r}\right)^{\frac{2}{4+\beta}} ~T_\mathrm{min},
\end{equation}
with the exponent $\beta$ of the opacity power law in the long wavelength regime and the temperature $T_\mathrm{min}$ at the inner rim of the computational domain at $R_\mathrm{min}$.
Usually, $\beta$ can be approximated as $\beta \approx 2$.
The exponent of the opacity power law in the long wavelength regime \citep{Draine:1984p118} is determined to be $\beta = 2.0508$ (see Fig.~\ref{fig:opacity}).
In the limiting case of gray and isotropic (i.e., non-irradiation) radiation fields, the local dust temperature is expected to follow
\begin{equation}
\label{eq:grayisotropic}
T(r) = \sqrt{\frac{R_\mathrm{min}}{r}} ~ T_\mathrm{min}.
\end{equation}
The analytical estimate \eqref{eq:Spitzer} by \citet{Spitzer:1978p776} and the gray, isotropic relation \eqref{eq:grayisotropic} are shown together with the simulation results for the most optically thin case in Fig.~\ref{fig:tau1e-1}.

As expected, 
the gray FLD approximation resembles the slope of the gray isotropic assumption,
while the frequency-dependent and gray hybrid methods resemble the Spitzer estimate for irradiated regions.
In this limiting case of very low optical depth, the frequency-dependent and gray irradiation routines give identical results.

\subsubsection{The most optically thick regime}
In general, configurations including regions of higher optical depth in otherwise optically thin environments denote the numerically most challenging problems.
Typically very accurate radiation transport methods such as RT and Monte-Carlo have to consume the majority of cpu time to reach a converged temperature in the optically thick regions.
While photon propagation in the optically thick regions alone is most easily described in the diffusion limit, the (flux-limited) diffusion approximation fails to account for shadowing effects in multi-dimensional problems due to the loss of angular information in its derivation.

The simulation results for the most optically thick case are shown in Fig.~\ref{fig:tau1e+6}.
The RT + gray FLD runs are in overall good agreement with the comparison results from RADMC.
The maximum deviation of 43\% corresponds to the maximum difference between the various radiation transport codes of the original benchmark test \citep{Pinte:2009p418}.
As expected in this limiting case of very high optical depth, the frequency-dependent and gray irradiation routines give identical results.

The gray FLD approximation fails to reproduce the shadow behind the optically thick inner disk rim.
In this method, the stellar radiative flux diffuses around this obstacle and heats the disk through the optically thin atmosphere at larger radii.
The resulting midplane temperature is up to a factor of 2.8 higher than in the comparison run.

\subsubsection{The regime of moderate optical depth}
The frequency-dependent and gray irradiation routines differ only in the case of moderate optical depth, shown in Fig.~\ref{fig:tau1e+2}.
In this intermediate regime, the disk's midplane is optically thick for the higher frequency part of the stellar irradiation spectrum, but optically thin in the long wavelength regime.
Therefore, only the frequency-dependent irradiation routine is able to resemble the heating of the irradiated midplane at larger radii correctly.
In the case of gray irradiation, the long wavelength flux is already absorbed at the inner disk rim and the temperature at larger radii follows the slope of the gray FLD method.

\subsection{Validity of approximations}
\subsubsection{The gray flux-limited diffusion approximation for thermal dust emission}
In all our approximate radiation transport simulations, the thermal (re-)emission of dust grains is computed in the gray FLD approximation.
This approximation reduces cpu time significantly (up to several orders of magnitude for a high-resolution 3D scheme) to allow the usage of the radiation transport method within magneto-hydrodynamics simulations.
If the stellar irradiation is accounted for within a frequency-dependent RT step, the results of the approximate radiation transport simulations are in accordance with the deviations given by the differences from the various high-level radiation transport methods from the original radiation benchmark tests \citep{Pascucci:2004p327,Pinte:2009p418}.
Hence, we conclude that the gray FLD approximation is an efficient and accurate method for solving the thermal dust (re-)emission of circumstellar disks.
In general, the FLD method suffers from the issues explained in Sect.~\ref{sect:FLD-star}.

\subsubsection{The gray approximation for the stellar irradiation}
The frequency-dependent treatment of the stellar irradiation spectrum improves the results for configurations of moderate optical depth (for the particular test case of a Sun-like star and $\tau_\mathrm{550nm} \sim 100$), in which the disk's midplane is optically thick for the high-frequency part of the stellar spectrum, while it is optically thin for  a substantial part of the long wavelength regime.
In the limiting cases of very low and very high optical depth, the gray irradiation is in agreement with the frequency-dependent runs.

\subsubsection{The flux-limited diffusion approximation for the stellar source term}
\label{sect:FLD-star}
In general,
treating the stellar source term within the FLD approximation involves two major problems.
First, the FLD approximation assumes a locally isotropic radiation field.
This assumption yields the wrong radial slope of the temperature even in highly optically thin regions, as depicted in Fig.~\ref{fig:tau1e-1}.
Dust grains in an optically thin region around a star absorb photons mostly from the direction towards the star.
Hence, this scenario is highly anisotropic. 

Second, in the derivation of the FLD approximation, the momentum expansion of the radiation transport equation is closed by relating the radiative flux to the gradient of the radiation energy.
This assumption prevents the FLD approximation from accounting for shadows behind radiative obstacles, such as the optically thick inner rim of circumstellar disks.
The radiative flux in FLD simulations tends to diffuse from the stellar surface around the optically thick inner disk rim and heats the disk midplane at larger radii to unrealistically high values.

Hence, we conclude that the FLD approximation results in large errors if used for (stellar) irradiation sources.
In (magneto-) hydrodynamic studies of accretion disks, the strong increase of the midplane temperature would result, for example, in an overestimate of the disk stability based on the Toomre criterion \citep{Toomre:1964p677}.

\subsubsection{The hybrid approximate radiation transport method}
In general,
the hybrid approximate radiation transport method using a frequency-dependent RT step for stellar irradiation and a gray FLD solver for thermal dust emission
results in high accuracy for the whole range of optical depths.
By using only a small fraction of the cpu time needed for the more sophisticated comparison code,
this method fully agrees with cases of optical depths $\tau_\mathrm{550nm} \le 100$ as well as with the high optical depth cases.
Only in the intermediate regime of a total optical depth of about $\tau_\mathrm{810nm} \sim 1000$ does the approximate method overestimate the temperature in the disk midplane behind the most optically thick inner rim by about 48\%, compared to only 20\% deviation of the various original benchmark codes in \citet{Pinte:2009p418}.

\section{Summary}
\begin{figure}[htbp]
\begin{center}
\includegraphics[width=0.5\textwidth]{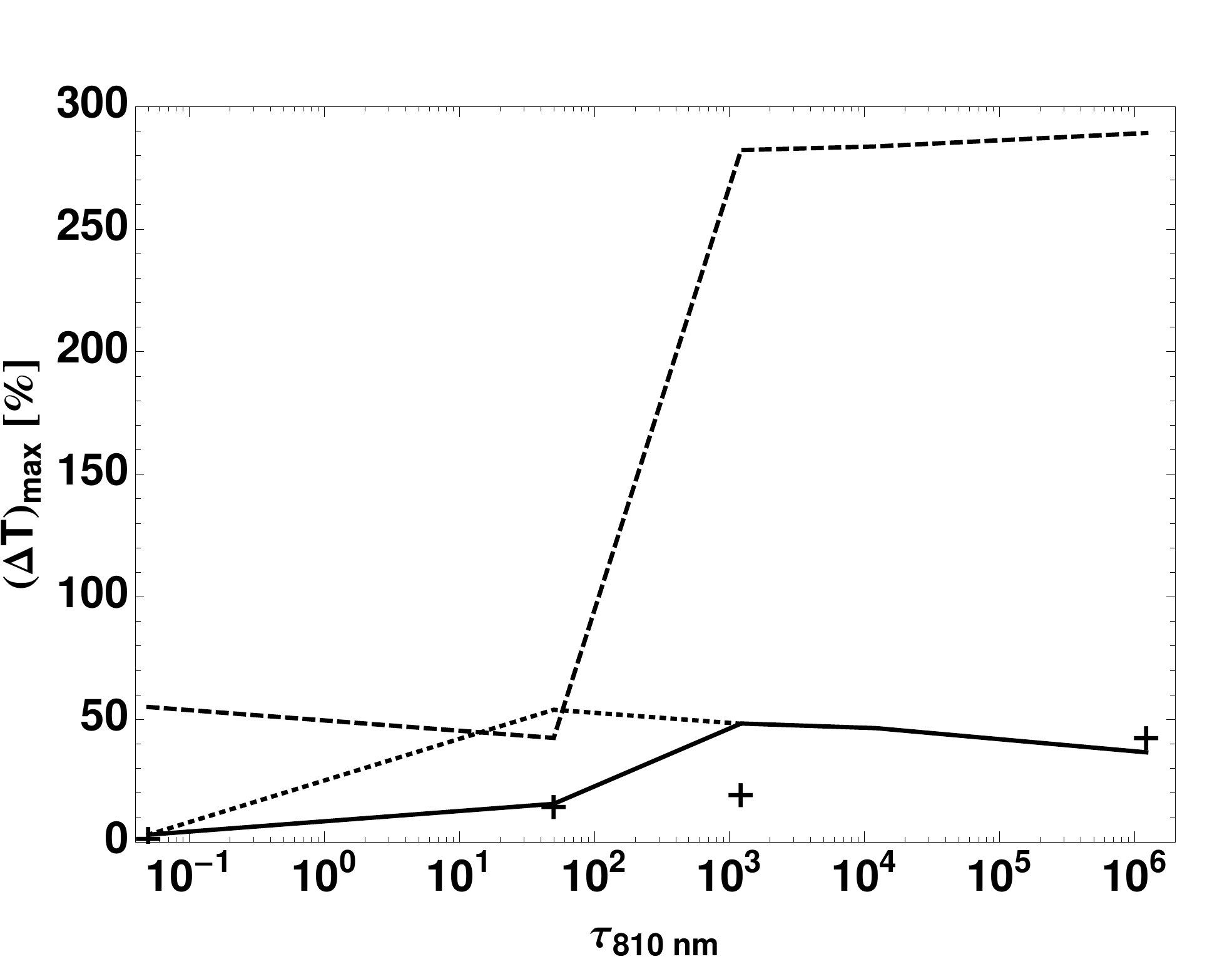}
\caption{
Maximum temperature deviation in the disk's midplane as function of the optical depth of the circumstellar disk.
The solid line denotes the results for the frequency-dependent RT + gray FLD method.
The dotted line denotes the results for the gray RT + gray FLD method.
The dashed line denotes the results for the gray FLD method.
The pluses ``+" denote the deviations of the results of the various high-level radiation transport codes used in the original benchmark tests of \citet{Pascucci:2004p327} and \citet{Pinte:2009p418}.
These values are read off Figs.~4 and 5 in \citet{Pascucci:2004p327} and Fig.~10 in \citet{Pinte:2009p418}.
}
\label{fig:deltaT}
\end{center}
\end{figure}

We checked the reliability of approximate radiation transport methods using flared disk configurations of standard radiation benchmark tests.
We compared the resulting temperature distributions for a wide range of optical depths from a fairly optically thin case ($\tau_\mathrm{550nm} = 0.1$) up to a highly optically thick case ($\tau_\mathrm{810nm} = 1.22\times10^{+6}$).
The outcome for the various methods and different regimes of optical depth are summarized in Fig.~\ref{fig:deltaT}.

Approximating the stellar irradiation spectrum by a gray Planck opacity is in accordance with a frequency-dependent treatment in the optically thin and thick limit.
For intermediate optical depths,
a gray approximation of the stellar irradiation spectrum yields a slightly hotter inner rim and a slightly cooler midplane of the disk at larger radii.

The gray FLD approximation fails to compute an appropriate temperature profile in all regimes of optical depth; the maximum deviations to the comparison runs are 50\% in the optically thin and up to 280\% in the optically thick limit. 
For low optical depth, the isotropic assumption within the FLD method yields a too steep decrease of the radial temperature slope. 
For higher optical depths, the FLD approximation does not render the shadow behind the optically thick inner rim of the circumstellar disk, yielding artificial heating at larger disk radii.
This issue of neglected shadows is expected to be even more problematic for increasing stellar luminosity.

The frequency-dependent RT + gray FLD approximation denotes a highly accurate treatment of the radiation field in the optically thin and thick regime.
Only in the intermediate regime of moderate optical depth $\tau_\mathrm{810nm} \sim 1000$ are the deviations of the radiation transport codes from the original benchmark tests by \citet{Pinte:2009p418} lower than for the hybrid approximate radiation transport method.
The high accuracy of the hybrid approximate radiation transport is attended by very low computational costs compared with the very expensive Monte-Carlo calculations.

Hence, we close this comparison study by pointing out that the frequency-dependent RT + gray FLD method is ideally suited for follow-up (magneto-)hydrodynamical studies of circumstellar accretion disks.

\begin{acknowledgements}
We thank Cornelis Dullemond for providing the RADMC reference simulation data and for fruitful discussions of approximate radiation transport solvers in the context of passively irradiated circumstellar disks.
Author R.~K.~acknowledges financial support from the {\em Deutsche Forschungsgemeinschaft} (DFG) via the collaborative research project SFB 881 ``The Milky Way System'' in sub-project B2. 
Both authors thank the DFG for support via the priority program SPP 1573 ``Physics of the ISM'' as well as the Baden-W\"{u}rttemberg Foundation for support via contract research (grant P- LS-SPII/18) via their program {\em Internationale Spitzenforschung II}.
\end{acknowledgements}

\bibliographystyle{aa}
\bibliography{Papers}

\begin{thebibliography}{47}
\expandafter\ifx\csname natexlab\endcsname\relax\def\natexlab#1{#1}\fi

\bibitem[{Alibert {et~al.}(2004)Alibert, Mordasini, \&
  Benz}]{Alibert:2004p17254}
Alibert, Y., Mordasini, C., \& Benz, W. 2004, A{\&}A, 417, L25

\bibitem[{Alibert {et~al.}(2011)Alibert, Mordasini, \&
  Benz}]{Alibert:2011p17315}
Alibert, Y., Mordasini, C., \& Benz, W. 2011, A{\&}A, 526, 63

\bibitem[{Alibert {et~al.}(2005)Alibert, Mordasini, Benz, \&
  Winisdoerffer}]{Alibert:2005p17281}
Alibert, Y., Mordasini, C., Benz, W., \& Winisdoerffer, C. 2005, A{\&}A, 434,
  343

\bibitem[{Balbus \& Hawley(1991)}]{Balbus:1991p816}
Balbus, S.~A. \& Hawley, J.~F. 1991, ApJ, 376, 214

\bibitem[{Bitsch {et~al.}(2012)Bitsch, Crida, Morbidelli, Kley, \&
  Dobbs-Dixon}]{Bitsch:2012p16070}
Bitsch, B., Crida, A., Morbidelli, A., Kley, W., \& Dobbs-Dixon, I. 2012,
  eprint arXiv, 1211, 6345

\bibitem[{Bjorkman \& Wood(2001)}]{Bjorkman:2001p645}
Bjorkman, J.~E. \& Wood, K. 2001, ApJ, 554, 615

\bibitem[{Boley {et~al.}(2007)Boley, Durisen, Nordlund, \&
  Lord}]{Boley:2007p642}
Boley, A.~C., Durisen, R.~H., Nordlund, {\AA}., \& Lord, J. 2007, ApJ, 665,
  1254

\bibitem[{Boss(1984)}]{Boss:1984p16962}
Boss, A.~P. 1984, MNRAS, 209, 543

\bibitem[{Chiang \& Goldreich(1997)}]{Chiang:1997p17115}
Chiang, E.~I. \& Goldreich, P. 1997, ApJ, 490, 368

\bibitem[{Draine \& Lee(1984)}]{Draine:1984p118}
Draine, B.~T. \& Lee, H.~M. 1984, ApJ, 285, 89

\bibitem[{Dullemond \& Dominik(2004{\natexlab{a}})}]{Dullemond:2004p523}
Dullemond, C.~P. \& Dominik, C. 2004{\natexlab{a}}, A{\&}A, 421, 1075

\bibitem[{Dullemond \& Dominik(2004{\natexlab{b}})}]{Dullemond:2004p1401}
Dullemond, C.~P. \& Dominik, C. 2004{\natexlab{b}}, Extrasolar Planets: Today
  and Tomorrow, 321, 361

\bibitem[{Dullemond \& Turolla(2000)}]{Dullemond:2000p686}
Dullemond, C.~P. \& Turolla, R. 2000, A{\&}A, 360, 1187

\bibitem[{Dullemond {et~al.}(2002)Dullemond, van Zadelhoff, \&
  Natta}]{Dullemond:2002p16279}
Dullemond, C.~P., van Zadelhoff, G.~J., \& Natta, A. 2002, A{\&}A, 389, 464

\bibitem[{Flaig {et~al.}(2012)Flaig, Ruoff, Kley, \&
  Kissmann}]{Flaig:2012p17017}
Flaig, M., Ruoff, P., Kley, W., \& Kissmann, R. 2012, MNRAS, 420, 2419

\bibitem[{Fouchet {et~al.}(2012)Fouchet, Alibert, Mordasini, \&
  Benz}]{Fouchet:2012p17303}
Fouchet, L., Alibert, Y., Mordasini, C., \& Benz, W. 2012, A{\&}A, 540, 107

\bibitem[{Gammie(2001)}]{Gammie:2001p25}
Gammie, C.~F. 2001, ApJ, 553, 174

\bibitem[{Godon \& Livio(2000)}]{Godon:2000p16974}
Godon, P. \& Livio, M. 2000, ApJ, 537, 396

\bibitem[{Harries(2011)}]{Harries:2011p1968}
Harries, T.~J. 2011, MNRAS, 416, 1500

\bibitem[{Harries {et~al.}(2012)Harries, Haworth, \&
  Acreman}]{Harries:2012p15058}
Harries, T.~J., Haworth, T.~J., \& Acreman, D. 2012, eprint arXiv, 1209, 1512

\bibitem[{Hawley \& Balbus(1991)}]{Hawley:1991p643}
Hawley, J.~F. \& Balbus, S.~A. 1991, ApJ, 376, 223

\bibitem[{Hirose \& Turner(2011)}]{Hirose:2011p17114}
Hirose, S. \& Turner, N.~J. 2011, ApJL, 732, L30

\bibitem[{Johnson \& Gammie(2005)}]{Johnson:2005p16981}
Johnson, B.~M. \& Gammie, C.~F. 2005, ApJ, 635, 149

\bibitem[{Kenyon \& Hartmann(1987)}]{Kenyon:1987p17217}
Kenyon, S.~J. \& Hartmann, L. 1987, ApJ, 323, 714

\bibitem[{Klahr \& Bodenheimer(2003)}]{Klahr:2003p779}
Klahr, H. \& Bodenheimer, P. 2003, ApJ, 582, 869

\bibitem[{Kley \& Lin(1992)}]{Kley:1992p560}
Kley, W. \& Lin, D. N.~C. 1992, ApJ, 397, 600

\bibitem[{Kley {et~al.}(1993{\natexlab{a}})Kley, Papaloizou, \&
  Lin}]{Kley:1993p480}
Kley, W., Papaloizou, J. C.~B., \& Lin, D. N.~C. 1993{\natexlab{a}}, ApJ, 416,
  679

\bibitem[{Kley {et~al.}(1993{\natexlab{b}})Kley, Papaloizou, \&
  Lin}]{Kley:1993p656}
Kley, W., Papaloizou, J. C.~B., \& Lin, D. N.~C. 1993{\natexlab{b}}, ApJ, 409,
  739

\bibitem[{Kratter \& Matzner(2006)}]{Kratter:2006p428}
Kratter, K.~M. \& Matzner, C.~D. 2006, MNRAS, 373, 1563

\bibitem[{Kratter {et~al.}(2008)Kratter, Matzner, \&
  Krumholz}]{Kratter:2008p196}
Kratter, K.~M., Matzner, C.~D., \& Krumholz, M.~R. 2008, ApJ, 681, 375

\bibitem[{Kuiper {et~al.}(2010{\natexlab{a}})Kuiper, Klahr, Beuther, \&
  Henning}]{Kuiper:2010p541}
Kuiper, R., Klahr, H., Beuther, H., \& Henning, T. 2010{\natexlab{a}}, ApJ,
  722, 1556

\bibitem[{Kuiper {et~al.}(2011)Kuiper, Klahr, Beuther, \&
  Henning}]{Kuiper:2011p349}
Kuiper, R., Klahr, H., Beuther, H., \& Henning, T. 2011, ApJ, 732, 20

\bibitem[{Kuiper {et~al.}(2012)Kuiper, Klahr, Beuther, \&
  Henning}]{Kuiper:2012p1151}
Kuiper, R., Klahr, H., Beuther, H., \& Henning, T. 2012, A{\&}A, 537, 122

\bibitem[{Kuiper {et~al.}(2010{\natexlab{b}})Kuiper, Klahr, Dullemond, Kley, \&
  Henning}]{Kuiper:2010p586}
Kuiper, R., Klahr, H., Dullemond, C.~P., Kley, W., \& Henning, T.
  2010{\natexlab{b}}, A{\&}A, 511, 81

\bibitem[{Kuiper \& Yorke(2013)}]{Kuiper:2013p17358}
Kuiper, R. \& Yorke, H.~W. 2013, ApJ, 763, 104

\bibitem[{Lin \& Bodenheimer(1982)}]{Lin:1982p16973}
Lin, D. N.~C. \& Bodenheimer, P. 1982, ApJ, 262, 768, a{\&}AA ID.
  AAA032.107.029

\bibitem[{Mordasini {et~al.}(2009{\natexlab{a}})Mordasini, Alibert, \&
  Benz}]{Mordasini:2009p17238}
Mordasini, C., Alibert, Y., \& Benz, W. 2009{\natexlab{a}}, A{\&}A, 501, 1139

\bibitem[{Mordasini {et~al.}(2012)Mordasini, Alibert, Benz, Klahr, \&
  Henning}]{Mordasini:2012p17258}
Mordasini, C., Alibert, Y., Benz, W., Klahr, H., \& Henning, T. 2012, A{\&}A,
  541, 97

\bibitem[{Mordasini {et~al.}(2009{\natexlab{b}})Mordasini, Alibert, Benz, \&
  Naef}]{Mordasini:2009p17253}
Mordasini, C., Alibert, Y., Benz, W., \& Naef, D. 2009{\natexlab{b}}, A{\&}A,
  501, 1161

\bibitem[{Owen {et~al.}(2012)Owen, Ercolano, \& Clarke}]{Owen:2012p16415}
Owen, J.~E., Ercolano, B., \& Clarke, C.~J. 2012, eprint arXiv, 1208, 6243

\bibitem[{Pascucci {et~al.}(2004)Pascucci, Wolf, Steinacker, Dullemond,
  Henning, Niccolini, Woitke, \& Lopez}]{Pascucci:2004p327}
Pascucci, I., Wolf, S., Steinacker, J., {et~al.} 2004, A{\&}A, 417, 793

\bibitem[{Pinte {et~al.}(2009)Pinte, Harries, Min, Watson, Dullemond, Woitke,
  M{\'e}nard, \& Dur{\'a}n-Rojas}]{Pinte:2009p418}
Pinte, C., Harries, T.~J., Min, M., {et~al.} 2009, A{\&}A, 498, 967

\bibitem[{Pollack {et~al.}(1996)Pollack, Hubickyj, Bodenheimer, Lissauer,
  Podolak, \& Greenzweig}]{Pollack:1996p17324}
Pollack, J.~B., Hubickyj, O., Bodenheimer, P., {et~al.} 1996, Icarus, 124, 62

\bibitem[{Spitzer(1978)}]{Spitzer:1978p776}
Spitzer, L. 1978, New York Wiley-Interscience, 333

\bibitem[{Toomre(1964)}]{Toomre:1964p677}
Toomre, A. 1964, ApJ, 139, 1217

\bibitem[{Umurhan \& Regev(2004)}]{Umurhan:2004p16978}
Umurhan, O.~M. \& Regev, O. 2004, A{\&}A, 427, 855

\bibitem[{Vaidya {et~al.}(2009)Vaidya, Fendt, \& Beuther}]{Vaidya:2009p436}
Vaidya, B., Fendt, C., \& Beuther, H. 2009, ApJ, 702, 567

\end{thebibliography}

\end{document}